\documentclass[twocolumn,reprint,amsmath,amssymb,footinbib,superscriptaddress,floatfix,aps,longbibliography,prl]{revtex4-2}
\pdfoutput=1

\usepackage[english]{babel}
\usepackage[dvipsnames]{xcolor}
\usepackage{graphicx}   
\usepackage{epstopdf}
\usepackage{amsfonts}
\usepackage{amstext,amsmath,amssymb}
\usepackage{lipsum}
\usepackage{rotating}
\usepackage{dcolumn}
\usepackage{array,multirow}
\usepackage[separate-uncertainty=true]{siunitx}
\DeclareSIUnit\stokes{St}
\DeclareSIUnit\gauss{G}
\usepackage[export]{adjustbox}
\usepackage{braket}
\usepackage[caption=false]{subfig}
\usepackage{xr}
\usepackage{ifpdf}
\usepackage{tikz}
\usetikzlibrary{arrows}
\usepackage{bm}
\usepackage{bbold}
\usepackage{orcidlink}
\usepackage{hyperref}
\begin{document}

\title{Optimally Driven Dressed Qubits}

\author{Alon Salhov \orcidlink{0000-0001-9111-2400}}
\email[Corresponding author: ]{alon.salhov@mail.huji.ac.il}
\affiliation{Racah Institute of Physics, The Hebrew University of Jerusalem, Jerusalem 9190401, Israel}
\author{Sagi Nechushtan}
\affiliation{Racah Institute of Physics, The Hebrew University of Jerusalem, Jerusalem 9190401, Israel}
\author{Alex Retzker}
\affiliation{Racah Institute of Physics, The Hebrew University of Jerusalem, Jerusalem 9190401, Israel}

\date{\today}

\begin{abstract}
The applicability and performance of qubits dressed by classical fields are limited because their control protocols give rise to an undesired counter-rotating term (CRT). This in turn forces operation in a regime where a (dressed) rotating-wave approximation (RWA) is valid, thereby restricting key aspects of their operation.
Here, using only a single coupling axis in the laboratory frame, we introduce a dressed-qubit control protocol that optimally removes the CRT, eliminating the need for the RWA and delivering substantial improvements in multiple performance metrics, including single-qubit gate speed, two-qubit gate fidelity, spectroscopic range, clock stability, and coherence preservation.
In addition, we provide a general parameterization together with a Floquet-based coherence-time expression, which elucidates the protocol's working principles and lowers the barrier to adoption.
Collectively, these advances position our scheme as the state-of-the-art strategy for qubit control, paving the way for a wider class of quantum technologies to be realized using dressed-qubit architectures.
\end{abstract}

\maketitle

\section{Introduction}\label{sec:Introduction}
Classical-field-dressed qubits are effective quantum systems that describe the behavior of a quantum two-level system (the bare qubit) interacting with a classical electromagnetic field. As in standard textbook treatments \cite{CohenTannoudji1998API}, for a linearly polarized, transition-inducing, monochromatic resonant field, the Hamiltonian
$
H = \frac{1}{2}\omega_0\sigma_z + \Omega_1\cos(\omega_0 t)\sigma_x,
$
with $\omega_0$ denoting the energy splitting of the bare qubit and $\Omega_1$ the Rabi frequency, can be transformed into a rotating frame (or interaction picture) with respect to $\frac{1}{2}\omega_0\sigma_z$. Under the bare rotating-wave approximation (RWA, $\Omega_1\ll\omega_0$, the qualifier “bare” will be explained shortly), this yields
$
H_I = \frac{1}{2}\Omega_1\sigma_x.
$
This resulting expression is the Hamiltonian of the dressed qubit, directly analogous to the field-free Hamiltonian of the bare qubit.
The related phenomena of Autler-Townes splitting \cite{miao2020universal}, AC-stark shift \cite{CohenTannoudji1998API} and Mollow triplet \cite{wang2021observationMollow}  are often described using dressed qubits.

For quantum technological applications such as information processing, sensing \cite{Degen2017RMP}, and atomic clocks, dressed qubits need to be well controlled. Several such driving protocols have been proposed, many of them explored in the context of coherence protection and dynamical decoupling \cite{Viola1999PRL,wang2020coherence}. These include a concatenated continuous driving scheme \cite{CaiNJP2012}, a protocol that uses continuous time-dependent detuning \cite{CohenFP2017}, and mixed dynamical decoupling in which the dressed driving field is pulsed \cite{Genov2019MDD}. Another method showed that destructive interference of cross-correlated amplitude noise can be further used to prolong coherence times \cite{PhysRevLett.132.223601}. These schemes were used in various experimental systems including solid-state defects \cite{PhysRevLett.132.223601,wang2020coherence,farfurnik2017experimental,CaiNJP2012,StarkNatComm2017,ramsay2023coherence,patrickson2024high,patrickson2025microwave,miao2020universal,kim2025suppression}, quantum dots \cite{laucht2016breaking,laucht2017dressed}, trapped ions \cite{PhysRevX2025Daniel,Multi-ionPRL2024}, and neutral-atom arrays \cite{wang2025individual}.

All of the schemes discussed above generate an additional counter-rotating term (CRT) in the dressed basis, and therefore rely on the dressed RWA. If we denote by $\Omega_2$ the Rabi frequency associated with the dressed-qubit control, then the dressed RWA requires $\Omega_2 \ll \Omega_1$. The situation is conceptually similar to the driven bare qubit with the bare RWA, where $\Omega_1 \ll \omega_0$. In practice, however, the two cases differ significantly: while the bare RWA is typically fulfilled because the relevant frequency scales stem from distinct physical origins, the dressed RWA is artificially enforced, as both frequencies are set by the applied control field. Importantly, the CRT limits performance in several key aspects including single-qubit speed, two-qubit gate fidelity, coherence times, sensing bandwidth, and clean coupling to external degrees of freedom. Moreover, the CRT can complicate the dynamics significantly \cite{laucht2016breaking,fuchs2009gigahertz}, and, in some schemes, "wastes" half the dressed drive amplitude, which further limits performance when amplitude, power, or energy constraints exist \cite{casanova2019modulated, aharon2019quantum,cao2020protecting,SMARThuang2024high,SMARThansen2022implementation,SMARTPhysRevA.104.062415,laucht2016breaking,laucht2017dressed}. 

Here, we propose a control protocol that enables arbitrary dynamic control of the dressed qubit Hamiltonian using only a \textit{single} coupling axis, i.e., without requiring any extra control hardware. We apply it to create rotating-frame circularly polarized fields, thereby eliminating the CRT in the dressed basis and obviating the need for the dressed RWA.
As a result, we discuss how dressed single-qubit gates can be made equally fast compared to the bare ones, while exhibiting enhanced robustness to control noise.
We then demonstrate numerically how our protocol improves performance in all aforementioned aspects, and analyze three recent experiments that stand to benefit from the resulting elimination of the dressed CRT.
First, we show a $\sim27$-fold improvement of the two-qubit gate fidelity in a trapped-ion quantum processor, compared to a recent scheme employed in \cite{PhysRevX2025Daniel}. The resulting infidelity is an order-of-magnitude below the quantum computing fault-tolerant threshold \cite{aharonov1997fault,shor1996fault}. 
Second, the scheme is used to extend the detection range of a quantum sensor by enhancing its sensivity $10$-fold, as compared to two recently demonstrated schemes \cite{hermann2024extending,PhysRevLett.132.223601}.
Third, we show a $5$-fold coherence time improvement under magic-angle constrained drive, when compared to a scheme recently used in a multi-ion atomic clock \cite{Multi-ionPRL2024}. 

Finally, by applying Floquet theory together with a scaling ansatz, we obtain a simple approximate expression for the coherence time ($T_2$) in terms of the control parameters and noise strengths, which serves as powerful tool for experimental design. Our analysis shows that the advantageous $T_2$ scaling of continuously decoupled dressed qubits (relative to bare qubits) can be maintained across a broad range of parameters, i.e., over a wide bandwidth. This broad robustness is crucial for realizing long coherence times when the accessible control parameter space is constrained, as is typically the case in the quantum computing, quantum sensing, and atomic clock scenarios we consider.

\section{Theoretical Framework}

\begin{figure*}[t]
  \centering
  \includegraphics[width=1\textwidth]{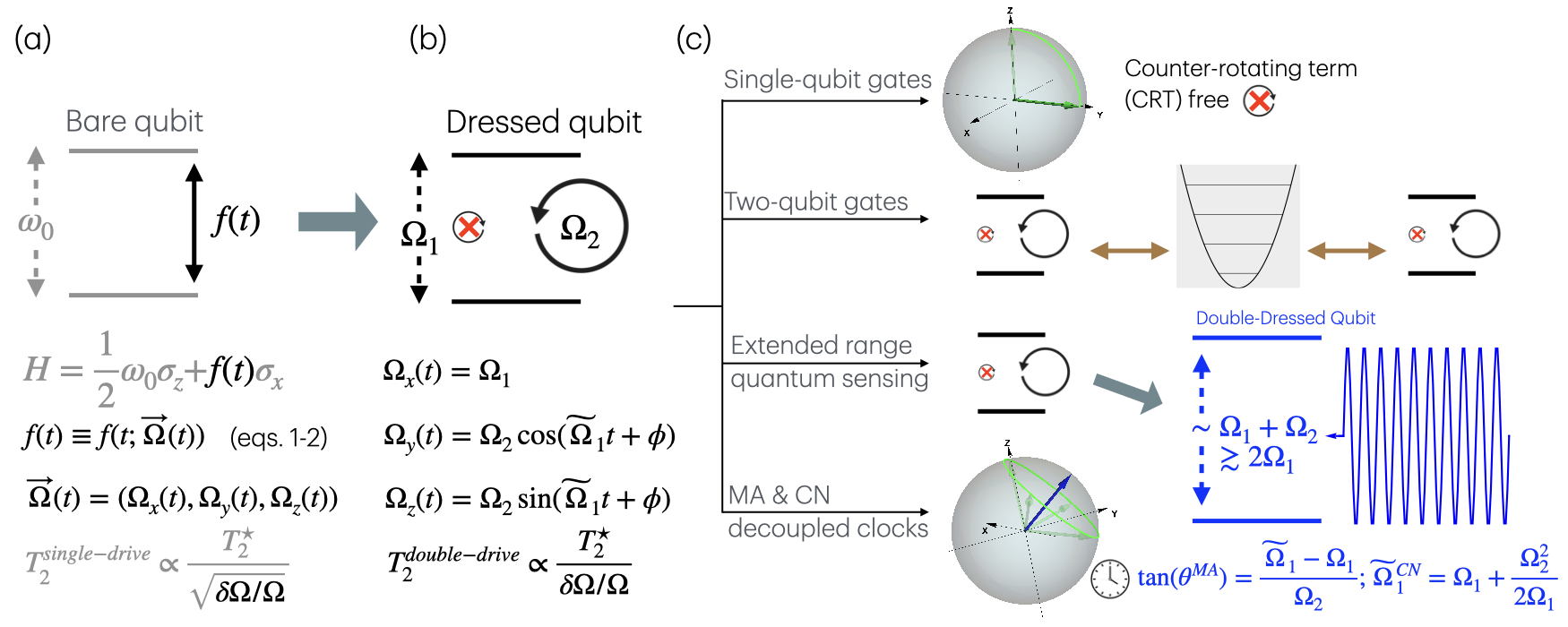}
\caption{Concept and applications of optimally driven dressed qubits. (a) A bare qubit is controlled to obtain an arbitrarily controlled dressed qubit (b), which can then be driven by a rotating-frame circularly polarized field that optimally removes the counter-rotating term (CRT). This leads to substantially extended coherence times over a broad bandwidth. (c) Our scheme improves performance in several key applications, namely, single-qubit gate speed, two-qubit gate fidelity, extended range high-sensitivity quantum sensing, and magic-angle (MA) $\And$ correlated-noise (CN) decoupled multi-ion frequency references. The performance is benchmarked against recent experiments reported in \cite{PhysRevX2025Daniel,hermann2024extending,Multi-ionPRL2024}. }
  \label{fig:concept}
\end{figure*}
\subsection{Drive Hamiltonian in Balanced Form}
We consider a driven two-level system with a Hamiltonian ($\hbar=1$, Fig. \ref{fig:concept}(a))
\begin{equation}\label{Eq:lab_frame_Hamiltonian}
    \begin{split}
       &H=
       \frac{1}{2}\omega_0\sigma_z + f(t) \sigma_x,
       ~\text{with}~ \\ 
       &f(t) = 
       \Omega_x(t) \cos{\left(\omega_0 t - \int_0^t \Omega_z(t')dt'\right)} \ -\\ 
       &\quad\quad\quad \Omega_y(t) \sin{\left(\omega_0 t - \int_0^t \Omega_z(t')dt'\right)},
    \end{split}
\end{equation}
where $\omega_0$ is the bare qubit energy gap, $f(t)$ is the control waveform, $\Omega_{x(y)}$ is the first (second) amplitude modulation, and $\Omega_z$ is a frequency modulation (i.e., time-dependent detuning or phase-modulation). Note that by using simple trigonometric identities, $f(t)$ can be recast into amplitude or phase/frequency modulation forms alone, and thus the specific experimental implementation should be determined by considering the control hardware capabilities and stability.

Transforming $H$ to a rotating frame, or interaction picture, with respect to $\frac{1}{2}(\omega_0 - \Omega_z(t)) \sigma_z$ and assuming $\omega_0$ is larger than any  rate or frequency in $\Omega_i(t)$ -- namely, under the bare RWA \cite{zeuch2020exact,Yudilevich_2023} -- we obtain
\begin{equation}\label{Eq:rotating_frame_Hamiltonian}
       H_I=
       \frac{1}{2}[\Omega_x(t)\sigma_x + \Omega_y(t)\sigma_y + \Omega_z(t)\sigma_z] = \frac{1}{2}\vec{\Omega}(t)\cdot\vec{\sigma},
\end{equation}
which we name the \textit{balanced-control} Hamiltonian (Fig. \ref{fig:concept}(b)).  Note that this arbitrary dressed qubit dynamic control of $H_I$, including modulation of its energy splitting, is achieved using a minimal \textit{single} lab-frame coupling axis. Namely, no additional control hardware is employed, such as secondary $y$ or $z$ polarized fields.

The aforementioned dressed qubit driving protocols generally begin by fixing $\Omega_x(t) = \Omega_1$, thereby defining an $x$-axis dressed qubit with energy splitting $\Omega_1$, and then implement dressed-qubit control through the $\Omega_y(t)$ or $\Omega_z(t)$ components (see SI for a survey). All of these methods are hindered by the dressed CRT and therefore must obey the \textit{control-only} dressed RWA, which, as we will show, degrade their performance. In the next section, we demonstrate how the balanced-control parameterization naturally suggests a remedy.

\subsection{CRT-free Dressed Qubit Control}

Although the existing drive schemes differ in several aspects (e.g., continuous vs. pulsed, detuning, modulation method, etc.), they all use fields that are linearly polarized in the dressed basis. Such driving fields can be decomposed into a sum of two circularly polarized fields, one is co-rotating with the precession induced by the dressed qubit energy gap and the other counter-rotating. It is the counter-rotating term (CRT) that can complicate the dynamics significantly and reduce performance.

Here, we introduce a scheme that drives the dressed qubit with a rotating-frame circularly polarized field, thereby optimally eliminating the CRT and the need for, and restrictions, of the dressed RWA. Crucially, it does not require any specialized resources or additional hardware beyond other dressed-control approaches—the lab-frame control still satisfies equation \ref{Eq:lab_frame_Hamiltonian} and uses only a single-axis, linearly polarized field. This contrasts with the bare qubit, where implementing circularly polarized driving demands phase-locked, two-axis control, as previously demonstrated for Nitrogen-Vacancy (NV) center spins with a cross-wire microwave antenna configuration \cite{london2014strong}, and in ultra-low-field NMR using two orthogonal coils \cite{shim2014strong}.

We proceed by fixing the dressed qubit energy splitting as $\Omega_x(t)=\Omega_1$ (as discussed above) and applying a transverse monochromatic dressed drive of the general form $\Omega_y(t)=\Omega_y\cos(\widetilde{\Omega}_1 t + \phi_y)$ and $ \Omega_z(t)=\Omega_z\sin(\widetilde{\Omega}_1 t + \phi_z)$. Transforming to the second interaction picture (of the dressed control), taken with respect to $\frac{1}{2}\widetilde{\Omega}_1\sigma_x$, yields a Hamiltonian containing a time-dependent contribution corresponding to the CRT. Nullifying the CRT implies $\Omega_y=\Omega_z\overset{\mathrm{def}}{=} \Omega_2$, and $\phi_y=\phi_z\overset{\mathrm{def}}{=} \phi$. 

As a result, the new scheme is obtained by setting - 
\begin{equation}\label{eq:new_scheme}
    \vec{\Omega}(t) =
    (\Omega_1,
    \Omega_2 \cos{(\widetilde{\Omega}_1 t + \phi)},
    \Omega_2 \sin{(\widetilde{\Omega}_1 t + \phi)})
\end{equation}
in eqs.~(\ref{Eq:lab_frame_Hamiltonian}-\ref{Eq:rotating_frame_Hamiltonian}); where $\Omega_1$ denotes the dressed qubit energy splitting (equivalently, the bare qubit Rabi frequency), $\Omega_2$ is the dressed qubit Rabi frequency (i.e., the second Rabi frequency), and $\widetilde{\Omega}_1$ represents the dressed qubit drive frequency (i.e., the modulation frequency).
Clearly, the $x$-basis dressed qubit is controlled by a $y-z$ plane circularly polarized field (Fig. \ref{fig:concept}(b)), and when transformed to the second interaction picture no CRT is present, obviating the need for the dressed RWA. The resulting exact Hamiltonian is $H_{II} = \frac{1}{2}[(\Omega_1-\widetilde{\Omega}_1)\sigma_x + \Omega_2(\cos(\phi)\sigma_y+\sin(\phi)\sigma_z)]$.

Observe that, as with conventional control of a bare qubit, the phase $\phi$ and detuning $\Omega_1-\widetilde{\Omega}_1$ control the dressed control axes.
Furthermore, observe that for existing methods the dressed-qubit Rabi frequency is only half of $\max|\Omega_{y/z}(t)|$, whereas in our protocol it is equal to the full value. This difference arises because alternative schemes effectively divert half of these modulation amplitudes into the counter-rotating terms (CRT).

In the next section, we showcase our scheme’s superior performance in several quantum technology applications (Fig. \ref{fig:concept}(c)). We demonstrate that, with our approach, dressed single-qubit gates can be executed as fast as bare qubit gates, while also exhibiting enhanced robustness against typically dominant control noise in this regime. We then consider an oscillator-mediated two-qubit gate and report more than an order-of-magnitude reduction in infidelity. Next, we show a tenfold increase in sensitivity for high–Rabi-frequency quantum sensing, and we conclude the section by demonstrating extended coherence times in the setting of multi-ion frequency-reference dynamical decoupling.

\section{Applications}

As has been shown for both bare qubits \cite{fuchs2009gigahertz,Yudilevich_2023} and dressed qubits \cite{laucht2016breaking}, the CRT can substantially complicate the dynamics, hindering experiments, data analysis, simulation, and theoretical understanding. The bare CRT seems unavoidable in the absence of (lab-frame) two axis control, and thus elaborate perturbative treatments were developed for the strong drive regime \cite{bloch1940magnetic,james2007effective,zeuch2020exact} (i.e., where the bare RWA is not valid). As we have seen, our protocol avoids this complication in the case of the dressed qubit.

Furthermore, in the context of single qubit gates, our scheme addresses two main disadvantages that arise in the comparison to bare qubits. 
First, with previous schemes, compliance with the dressed RWA means slower dressed operations. Specifically, the dressed CRT adds undesirable dynamics that induce gate infidelity \cite{BowdreyPhLett2002,nielsen2002simple} which depends on the amplitude ratio $\frac{\Omega_2}{\Omega_1}$. Recognizing that its inverse is the bare-to-dressed gate overhead time, sub-threshold infidelity typically requires an order-of-magnitude slowdown (see SI for details). Although this can be partially mitigated by using analytical or numerical techniques (e.g. incorporating the Bloch-Siegert shift \cite{bloch1940magnetic, aharon2019quantum} and competing terms \cite{zeuch2020exact}, pulse shaping \cite{zwanenburg2025single}, composite pulses \cite{torosov2021coherent,Genov2014}, and optimal control \cite{scheuer2014precise}), our scheme avoids it completely. Thus, dressed qubit gates using our scheme are equally fast as the bare qubit ones, which is hard to attain using the mentioned techniques in a simple and unified way, for all gates simultaneously (see SI).

The second issue relates to the effect of control noise on infidelity. Since this noise is often proportional to the control amplitude and fast gates require a high Rabi frequency, the control instability often becomes the dominant source of infidelity \cite{CaiNJP2012}. This means, in turn, that as far as amplitude-proportional, and control-noise-dominated, infidelity is concerned, dressed qubits bear no advantage over bare qubits. Put differently, the Q-factor is the same due to the proportional nature of the noise. 
When the phase stability of the control field surpasses its amplitude stability, as is often the case, this problem can be partially avoided by employing the time-dependent detuning approach \cite{CohenFP2017}. However, this method remains constrained by the CRT, which in turn enforces slow gate operation.  
A more comprehensive resolution—applicable also to fast gates—can be achieved by integrating our new scheme with the ideas in \cite{PhysRevLett.132.223601}, where it was demonstrated that cross-correlated noise can be made to destructively interfere, thereby having the potential to mitigate the noise proportionality problem by canceling it to first order in suitable scenarios.
This approach relies on the high cross-correlated fluctuations that were shown to exist for the dressed qubit ($\delta\Omega_1\approx\delta\Omega_2$), but they are not expected for the bare qubit ($\delta\omega_0\not\approx\delta\Omega_1$), due to their different origins. 

In the following sections, we demonstrate our protocol enhanced performance through three different applications, namely, reduced infidelity of a two-qubit gate, extended detection range of a quantum sensor, and prolonged coherence time in the context of atomic clocks.

\subsection{High Fidelity Two-Qubit Gate}

The time-dependent detuning scheme \cite{CohenFP2017}, also known as the phase-modulated (PM) drive and previously mentioned in the main text and described in the SI, was recently employed to implement a fast, rf-controlled two-qubit gate in a trapped-ion quantum processor that couples qubits via a static magnetic field gradient \cite{PhysRevX2025Daniel}. We begin by briefly outlining the method and the resulting infidelity, and then show that our new scheme achieves an approximately 27-fold reduction in the infidelity, bringing it an order-of-magnitude below the quantum error correction threshold.

The system of two ionic qubits trapped in a 1D harmonic potential subject to a linear magnetic field gradient is modeled with the lab frame Hamiltonian \cite{PhysRevX2025Daniel} 

\small\begin{equation}\label{eq:ion_full}
    H = \nu b^\dagger b + \sum_{j=1,2}\left(\frac{\omega_0^{(j)}}{2}\sigma_z^{(j)} + \frac{\eta\nu}{2}\sigma_z^{(j)}(b+b^\dagger) +H_D^{(j)}\right),
\end{equation}\normalsize
where $\nu$, $b^\dagger$ and $b$ are the motional mode's frequency, raising and lowering operators, respectively, $\omega_0^{(j)}$ and $\sigma_i^{(j)}$ are the transition frequency and Pauli operators of the $j^{th}$-ion two hyperfine states, $\eta$ is the effective Lamb-Dicke parameter, and $H_D^{(j)}$ is the drive Hamiltonian on the $j^{th}$-ion. The coupling to the stretch mode is neglected because it is far off resonance \cite{PhysRevX2025Daniel}.

The previous study employed the PM scheme \cite{PhysRevX2025Daniel}, setting
\begin{equation}\label{eq:ion_drive}
    H_{D-PM}^{(j)} = \Omega_1\cos\left(\omega_0^{(j)}t+\frac{\Omega_2}{\Omega_1}\sin(\Omega_1t)\right)\sigma_x^{(j)},
\end{equation}
to implement a fast two-qubit gate, where the interaction is mediated by the ions’ vibrational motion in the trap and the magnetic-field gradient. This is accomplished by choosing $\Omega_1=\nu-\eta\nu$ to induce a Mølmer–Sørensen–type gate at $t_g\approx\frac{2\pi}{\eta\nu}$; the modulation $\nu-\Omega_1\ll\Omega_2\ll\nu+\Omega_1$ decouples the dominant unwanted interaction terms, leaving a residual interaction corresponding to a quantum Stark shift, at the cost of introducing a CRT ($\Omega_2\ll\Omega_1$) \cite{sorensen1999quantum,sorensen2000entanglement,tan2013demonstration}.

This drive Hamiltonian corresponds to $\vec{\Omega}(t) =(\Omega_1,0,-\Omega_2 \cos{(\Omega_1 t))}$ in eq.~\eqref{Eq:rotating_frame_Hamiltonian}. Note that compared to the SI description of the PM scheme, this notation corresponds to $\Omega_2$ being \textit{twice} the second Rabi frequency; nevertheless, we adopt this notation in this section to make the comparison with previous work \cite{PhysRevX2025Daniel} straightforward. We shall consider all parameters of equations \ref{eq:ion_full}-\ref{eq:ion_drive} fixed to their values in \cite{PhysRevX2025Daniel}, except $\Omega_2$, which we scan to minimize the infidelity (see the SI for definition, parameter values, and simulation details).

\begin{figure}
    \centering
    \includegraphics[width=1\linewidth]{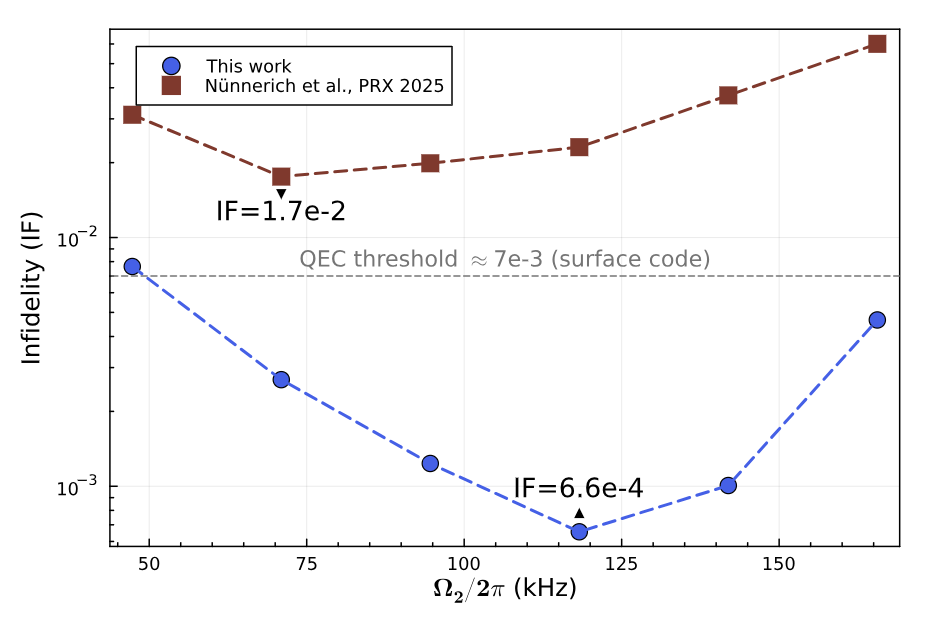}
    \caption{Two-qubit gate infidelity as a function of $\Omega_2$. Our new scheme using the optimal dressed drive (blue circles), at its optimal point, improves the infidelity by a factor of $\sim27$ compared to the optimum of the previous scheme in \cite{PhysRevX2025Daniel} which uses the phase-modulated (PM) drive (brown squares). Furthermore, the new optimal infidelity is below the QEC threshold (gray line), by an order-of-magnitude.}
    \label{fig:Fidelity2QG}
\end{figure}

Figure \ref{fig:Fidelity2QG} shows that for the PM scheme (brown squares), the simulated infidelity reaches a minimum of $1.7\times10^{-2}$ (annotated), insufficient for fault-tolerant quantum computation as it is above the quantum error correction (QEC) threshold ($\approx7\times10^{-3}$ for the surface code, gray line). This is due to an interplay of the two previously mentioned effects; for low $\Omega_2$ values ($\sim\nu-\Omega_1$), the quantum Stark shift \cite{PhysRevX2025Daniel}, a state-dependent frequency shift that perturbs the entangling operation, is poorly decoupled (the motional mode starts in a thermal state with $\bar n = 0.6$ phonons, as in \cite{PhysRevX2025Daniel}, see SI for details). For high $\Omega_2$ values on the other hand ($\sim\Omega_1$), it is the CRT of the drive that degrades the fidelity. The optimal $\Omega_{2-PM} \approx 2\pi \times 71$ kHz is where their combined effect is minimized and, as such, is the value used in the experiment \cite{PhysRevX2025Daniel}. The experimentally measured infidelity is in close agreement with the simulated value of approximately $1.7\times10^{-2}$.

Our new protocol is achieved by setting 
\small
\begin{equation}
\begin{aligned}
    H_{D-N}^{(j)} ={}&
    \Big[\Omega_1\cos\Big(\omega_0^{(j)} t + \frac{\Omega_2}{\Omega_1}\sin(\Omega_1 t)\Big)-\\
    &\ \ 
    \Omega_2  \sin\Big(\omega_0^{(j)} t + \frac{\Omega_2}{\Omega_1}\sin(\Omega_1 t)\Big)
    \sin(\Omega_1t) \Big] \, \sigma_x^{(j)},
\end{aligned}\label{eq:new_ion_drive}
\end{equation}
\normalsize
which corresponds to Eq. \ref{eq:new_scheme}, on resonance ($\widetilde{\Omega}_1 \to \Omega_1$), and $\phi=-\frac{\pi}{2}$ to match the conventions of the previous work \cite{PhysRevX2025Daniel}, namely, $\vec{\Omega}(t) =(\Omega_1,\Omega_2 \sin{(\Omega_1 t)},-\Omega_2 \cos{(\Omega_1 t)})$. Here, the CRT is eliminated and, as a result, the infidelity is significantly reduced
(blue circles in Fig. \ref{fig:Fidelity2QG}). Higher $\Omega_2$ values are thus allowed and better decouple the quantum Stark shift, until the second resonance at $\Omega_2 \sim \nu+\Omega_1$ is approached. Under these conditions, we achieve an optimal infidelity of $6.6\times10^{-4}$ for the new protocol, representing a $\sim 27$-fold improvement over the PM scheme. Furthermore, unlike for the PM  case, the resulting infidelity is below the QEC threshold, by an order-of-magnitude.

The optimal infidelity can be further reduced by additional initial cooling of the motional mode from the $\bar n=0.6$ thermal state (fig. \ref{fig:Fidelity2QG}) down to its ground state ($\bar n=0$), in which case the optimal infidelity reaches $1.2\times10^{-4}$ (for the same optimal $\Omega_2$ value). The residual infidelity at the optimal point scales as $\eta^2$, as follows from dimensional analysis and the findings reported in \cite{PhysRevX2025Daniel}. Future work could aim at still lower infidelities. For instance, by optimizing the trap frequency and magnetic field gradient, prolonging the gate duration, working with the stretch mode, or applying pulse-shaping techniques to achieve high-fidelity transfer from the (classical-field-)dressed states to the motional-mode-dressed states. At this point, however, additional sources of infidelity are likely to become relevant. In particular, one must consider heating of the motional mode during the gate, residual coupling to the stretch mode, and any remaining control noise \cite{PhysRevX2025Daniel}; the latter could be further mitigated by employing the methods of \cite{PhysRevLett.132.223601}.

\subsection{Extended Range Quantum Sensing}
A qubit sensor can efficiently detect signals that are resonant with its transition \cite{Degen2017RMP}.
By continuously driving a qubit, additional dressed transitions can be created and tuned at the Rabi frequency scale ($\sim\Omega_1$) \cite{PhysRevLett.132.223601,Genov2019MDD,CaiNJP2012} to extend its detection range, provided that the transition sensitivity is sufficiently high \cite{hermann2024extending}.
The sensitivity, which is a central figure-of-merit for quantum sensors, depends on several factors, two of them strongly influenced by the qubit's drive.
First, the coherence time $T_2$, which is prolonged when the control efficiently decouples noise, and second, the effective coupling (or transduction parameter) $\gamma \alpha$, since the bare coupling $\gamma$ is modified (typically attenuated) by a control-dependent factor $\alpha$ under the control-induced dynamics \cite{PhysRevLett.132.223601,patrickson2025microwave,StarkNatComm2017}.

The typical dependence of the sensitivity on these parameters follows $\eta = \frac{r}{\gamma C}\frac{1}{\alpha\sqrt{T_2}}$, where $r$ is a numerical prefactor and $C$ is a dimensionless constant quantifying readout efficiency \cite{Degen2017RMP}.
As a specific example, we can consider single NV-center parameters $r=\sqrt{8e}$, $\gamma=2\pi\times28$ Hz/nT, and $C=C_0\sqrt{N_{ph}}=0.24\sqrt{0.15}\approx0.093$, where $C_0$ is the initial NV readout contrast and $N_{ph}$ is the average number of photons collected per measurement run, as demonstrated in \cite{PhysRevLett.132.223601}. Note that $r,\ C$ and $\gamma$ are typically independent of the control protocol. To focus on the relevant terms - we define the sensitivity gain $\eta_{ref}/\eta$, with respect to a reference protocol.

To determine the coherence time, we use a noise model consisting of two contributions. First, the environment induces fluctuations to the qubit's energy gap, modeled as $\omega_0\to\omega_0+\delta$ (in Eq. \ref{Eq:lab_frame_Hamiltonian}), where $\delta \sim \mathcal{N}\!\left(0,\;\frac{2}{T_2^{\ast2}}\right)$ ($T_2^\ast$ being the bare qubit's Ramsey dephasing time). Second, control noise in the form of fractional amplitude fluctuations, modeled as $f(t)\to f(t)(1+\epsilon)$, where $\epsilon \sim \mathcal{N}\!\left(0,\;\sigma_\epsilon^2\right)$ (see SI for the definition and simulation details of $T_2$). 
The sensor's detection range is limited, therefore, since large Rabi frequencies necessarily imply the presence of significant control noise, affecting the coherence time and sensitivity. This regime is of special interest in scenarios requiring limited power, e.g. in living cells \cite{cao2020protecting}, and for magnetic resonance signals at high external magnetic fields. We set $T_2^\ast=1\mu s$ \cite{kitamura2025robust,tyler2025extended} and $\sigma_\epsilon =0.5\%$ \cite{PhysRevLett.132.223601, Genov2019MDD}, and note that although these are representative parameters for standard setups, our findings also generalize to setups with different noise parameters. 

To demonstrate our protocol's extended detection range, we compare it to two recently demonstrated sensing protocols.
First, sensing based on the Hartmann-Hahn (HH) condition \cite{hartmann1962nuclear} was demonstrated with an ensemble of Nitrogen Vacancy (NV) center spins in diamond \cite{hermann2024extending} to significantly extend the detection range in the high Rabi regime.
This protocol is based on a single drive $\vec{\Omega}(t) =(\Omega_1,0,0)$ and a resonance, or matching condition, $\Omega_1 \stackrel{!}{=} \omega_s $ to a signal of frequency $\omega_s$.
The second protocol we compare with, by some of the authors, utilized destructive interference of correlated control noise by setting $\vec{\Omega}(t) =(\Omega_1,2\Omega_2 \cos(\widetilde{\Omega}_1),0)$ where $\widetilde{\Omega}_1 \to \Omega_1+(c+\frac{1}{4})\frac{\Omega_2^2}{\Omega_1}$. Here, c is the cross-correlation parameter and equals $1$ for our noise model (note that, unlike the original work, we use this protocol for the detection of a Rabi-scale signal, accomplished by the matching condition discussed in the following)\cite{PhysRevLett.132.223601}.

Our new protocol is achieved by setting 
\begin{subequations}\label{sensing_scheme_correlated}
\begin{align}
    & \vec{\Omega}(t) =(\Omega_1,\Omega_2 \cos{(\widetilde{\Omega}_1 t)},\Omega_2 \sin{(\widetilde{\Omega}_1 t)});
    \\& \text{where } 
    \widetilde{\Omega}_1\to \Omega_1 + \frac{1}{2}\frac{\Omega_2^2}{\Omega_1},
\end{align}
\end{subequations}
and the detuning prefactor $\frac{1}{2}$ arises from the energy gap variance-minimization consideration, as in  \cite{PhysRevLett.132.223601}, for this noise model (i.e., for $c=1$). The analog of the Hartmann-Hahn matching condition for both two-tone protocols is $\widetilde{\Omega}_1+m\sqrt{(\widetilde{\Omega}_1-\Omega_1)^2+\Omega_2}\stackrel{!}{=}\omega_s$, where $m\in\{-1,0,1\}$ \cite{CaiNJP2012,aharon2019quantum,PhysRevLett.132.223601}. Since we are interested in the upper limit for the detection range, we proceed with condition $m=1$, and note that for $\Omega_2\to0$ we have $\widetilde{\Omega}_1\to \Omega_1$ and $\Omega_1\stackrel{!}{=}\omega_s$, namely, the 3 schemes coalesce (for $\Omega_2>0$ we assume that the detuning between the matching conditions is larger than the signal strength, i.e., we are interested in the detection limit for weak signals).

Sensing of a target signal with $\omega_s = 2\pi\times 200 \text{ MHz } \sim( T_2^\ast \sigma_\epsilon)^{-1}$ using the HH-condition is expected to deliver limited sensitivity, since for such high Rabi frequencies the coherence time scales as $\sim (\Omega_1\sigma_\epsilon)^{-1}\sim 1 \mu s$ \cite{CaiNJP2012}. Indeed, recent implementations of this scheme, up to $\sim 85$ MHz, demonstrated a reduced sensitivity as $\Omega_1$ increases \cite{hermann2024extending}. Although higher Rabi frequencies are technically achievable (for example, $\sim 500$ MHz \cite{fuchs2009gigahertz}), they offer limited benefit for quantum sensing if they are not accompanied by long coherence times.

\begin{figure}
    \centering
    \includegraphics[width=1\linewidth]{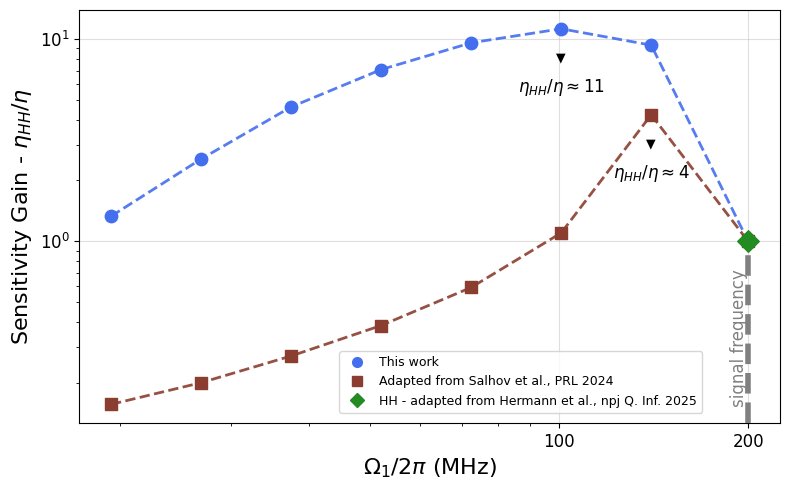}
    \caption{Sensitivity gain of our new protocol (blue circles) compared to the Hartmann-Hahn (HH) sensing protocol (green diamond) \cite{hartmann1962nuclear,hermann2024extending}. At the optimum ($\Omega_1/2\pi=100$ MHz), the gain exceeds an order of magnitude, which translates to over two orders-of-magnitude reduction in integration time for a given signal-to-noise ratio. The sensing scheme in \cite{PhysRevLett.132.223601}, adapted to the signal matching condition (see main text), provides a more modest gain (brown squares) despite accounting for the Bloch-Siegert shift. This demonstrates that the residual effects of the counter-rotating terms compromise the gain significantly. Finally, the gain provided by the new scheme is more robust, as it spans across a wider $\Omega_1$ range.}
    \label{fig:sensing}
\end{figure}

Figure \ref{fig:sensing} demonstrates that our new protocol (blue circles) enhances the sensitivity by more than an order of magnitude, compared to HH sensing (green diamond) \cite{hermann2024extending,hartmann1962nuclear}. At its optimal point, $\Omega_1/2\pi = 100$ MHz, the amplitude ratio is $\Omega_2/\Omega_1 = 0.7$, $\alpha = 0.17$ (cf. $\alpha_{HH}=0.5$), and $T_2= 254 \mu$s. Thus, sensitivity gain is achieved despite the strong attenuation factor and at a regime of a large amplitude ratio, through a long coherence time. Comparison to the second protocol (brown squares, adapted to the matching condition), reveals that while some advantage is conferred through its coherence protection properties, and despite the fact that the Bloch-Siegert shift is already incorporated -- the CRT still degrade performance, limiting the enhancement to a factor of 4. Finally, it is worth noting that, under our new protocol, the absence of the CRT leads to a simpler and more transparent time evolution, which simplifies measurement and data analysis.

\subsection{Stabilizing Atomic Clocks}
The second-order concatenated continuous driving scheme \cite{CaiNJP2012}, also known as the double-drive and previously mentioned in the main text and described in the SI, was recently employed (together with an off-resonant variant) to realize a stable multi-ion frequency reference based on $^{40}$Ca$^+$ \cite{Multi-ionPRL2024}.
The coherence time of the trapped-ions' interrogated optical transition was prolonged by suppression of its leading frequency shifts, namely, Zeeman shifts, driving amplitude shifts, and the electric quadrupole shift (QPS).

The Zeeman manifolds of the ground $^2S_{1/2}$ and excited $^2D_{5/2}$ levels of the optical transition are split by a static external magnetic field and dressed by control fields obeying $\vec{\Omega}^i(t) = (\Omega^i_1,2\Omega^i_2 \cos(\widetilde{\Omega}^i_1t),0)$, where $i \in \{S, D\}$. For the $S$-level, the standard double-drive is applied by setting $\widetilde{\Omega}^S_1 =\Omega^S_1$. For the $D$-level, which is affected by the QPS, stabilization is achieved by detuning  $\widetilde{\Omega}^D_1$ from resonance according to the magic-angle condition $(\widetilde{\Omega}^D_1-\Omega^D_1)/{\Omega^D_2} = 1/\sqrt2$. The magic-angle detuning suppresses the QPS at the expanse of reduced control noise decoupling efficiency. The remaining inaccuracy of this scheme arises from amplitude fluctuations $(\epsilon)$ as well as magnetic field noise $(\delta)$ \cite{Multi-ionPRL2024}. 

In the experiment, the amplitude ratio that was used to balance these contributions, while satisfying the dressed RWA, is $\Omega^D_2/\Omega^D_1\approx0.03$.
The same ratio was found numerically in a preceding theoretical study \cite{aharon2019robust}.
Note that both studies used a convention where $\Omega^i_2$ equals $2\times$ the dressed Rabi frequency. In this section, we shall follow the $\Omega^i_2$ equals ($1\times$) dressed Rabi convention, as described in the theory section. 

As discussed in earlier sections, it has been demonstrated \cite{PhysRevLett.132.223601} that when amplitude fluctuations are cross-correlated, the double-drive decouples noise most efficiently under an optimal detuning given by $\widetilde{\Omega}^i_1 \to \Omega^i_1+(c+\frac{1}{4})\frac{{\Omega^i_2}^2}{\Omega^i_1}$, where $c$ denotes the cross-correlation parameter (hereafter, we set $c=1$, as in the previous section). Indeed, for each level, both drives share control hardware and are therefore likely to exhibit correlated noise \cite{PelzerSchmidtPrivateComm}.

These findings suggest a potential to improve stability by operating at this optimal detuning.
This can be immediately applied to the $S$-level without requiring any additional changes to the drive parameters. 
For the $D$-level, in contrast, satisfying the optimal detuning and magic-angle conditions \textit{simultaneously} constrains the amplitude ratio for the double-drive to $\Omega^D_2/\Omega^D_1 = \sqrt{8}/5\approx0.6$, at which point the CRT is non-negligible and is expected to degrade stability. For our new protocol, the optimal detuning reads $\widetilde{\Omega}^D_1 \to \Omega^D_1+\frac{1}{2}\frac{{\Omega^D_2}^{2}}{\Omega^D_1}$ and the corresponding amplitude ratio is even higher, namely $\Omega^D_2/\Omega^D_1 = \sqrt{2}\approx1.4$, yet the lack of CRT makes it possible to operate in this unusual (ratio $> 1$) amplitude regime.

\begin{figure}
    \centering
    \includegraphics[width=1\linewidth]{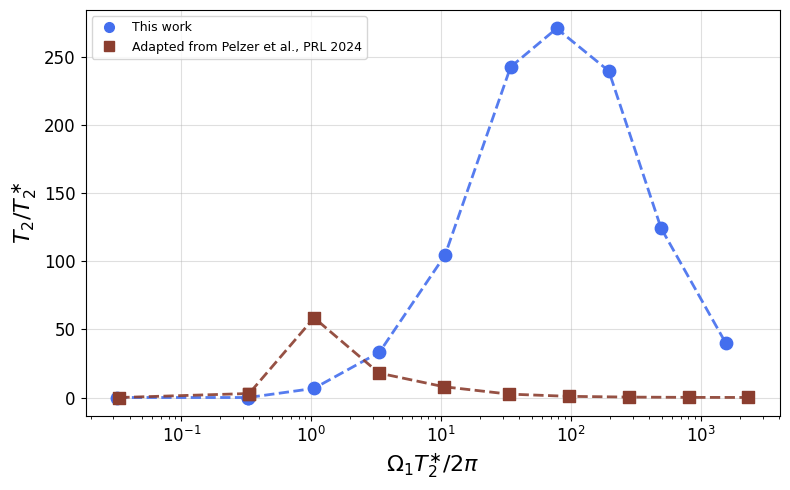}
    \caption{Zeeman coherence times as a function of $\Omega_1$ ($T_2^\ast$-scaled). At the optima, our new protocol (blue circles) extends coherence time by a factor of $\sim5$ compared to the magic-angle detuned double-drive used in \cite{Multi-ionPRL2024} (brown squares). Scalings are expected to depend on control noise strength $\sigma_\epsilon$ ($=0.5\%$ here, see discussion around Eq.~\eqref{eq:T2app}).
    }
    \label{fig:clock}
\end{figure}

To compare the performance of our new protocol to the magic-angle detuned double-drive in \cite{Multi-ionPRL2024}, we calculate the coherence time $T_2$ of a simplified system governed by Eq. \eqref{Eq:rotating_frame_Hamiltonian} and the same noise model as in the previous section ($\epsilon,   \delta$). We scale the time by $T_2^\ast (= \sqrt2/\sigma_\delta)$ and keep $\sigma_\epsilon=0.5\%$. For both protocols, we scan $\Omega_1$ to optimize $T_2$ (hereafter, we drop the $D$ superscript). As we previously mentioned, for the magic-angle detuned double-drive $\Omega_2$ and $\widetilde{\Omega}_1$ are set according to $\Omega_2/\Omega_1=0.03\%$ and $(\widetilde{\Omega}_1-\Omega_1)/{\Omega_2}=\frac{1}{\sqrt2}$, while for our new protocol they are set according to $\Omega_2/\Omega_1=\sqrt{2}$ and $(\widetilde{\Omega}_1-\Omega_1)/{\Omega_2} = \frac{1}{\sqrt2}$, satisfying the magic-angle and correlated noise conditions simultaneously. As Fig. \ref{fig:clock} demonstrates, our new protocol prolongs the coherence time by a factor of $\sim5$, compared to the scheme previously used. For context, note that the coherence time measured in \cite{Multi-ionPRL2024}, if enhanced by this factor, would exceed the natural decay time of the ions \cite{Multi-ionPRL2024,kreuter2005experimental}.

\section[Coherence Optimization via Floquet Theory]{Coherence Optimization via \\ Floquet Theory}
A key characteristic of qubit control protocols is dynamical decoupling, which serves to extend the qubit’s coherence time and mitigate noise-induced information loss. To discover how the coherence time depends on the control parameters, we analyze the dynamics governed by the Hamiltonian in Eq. \eqref{Eq:lab_frame_Hamiltonian}, incorporating our protocol in Eq. \eqref{eq:new_scheme}, while accounting for bare energy gap and control amplitude fluctuations.

These fluctuations, which often limit performance, are modeled by the substitutions $\omega_0 \rightarrow \omega_0 + \delta$ and $f(t)\rightarrow f(t)(1+\epsilon)$, with $\delta \sim \mathcal{N}(0,\sigma_\delta^2)$, $\epsilon \sim \ \mathcal{N}(0,\,\sigma_\epsilon^2)$ and $\sigma_{\delta} = \frac{\sqrt2}{T_2^\ast}$ ($T_2^\ast$ being the bare qubit's Ramsey dephasing time). Note that this assumes a perfect correlation between the amplitude fluctuations of the two amplitudes $\delta\Omega_{x/y}$, as recently demonstrated in \cite{PhysRevLett.132.223601}, and the high stability of the drive's frequency and phase control. The noise terms are modeled as random variables, not processes, effectively assuming that their correlation time is longer than the time of a single experiment. Note, however, that decoupling is often effective for shorter correlation times, as long as they are longer than the typical control timescale, or Rabi period, as demonstrated in \cite{PhysRevLett.132.223601}.
Finally, we take advantage of the freedom in selecting our time units and rescale time by $T_2^\ast$.

We proceed by transforming to the doubly rotating frame, first with respect to $\frac{1}{2}(\omega_0 - \Omega_2 \sin{(\widetilde{\Omega}_1 t)}) \sigma_z$, and then with respect to $\frac{1}{2}\widetilde{\Omega}_1\sigma_x$. Assuming the bare RWA we obtain

\begin{subequations}\label{eq:pmdd_subequations}
\begin{align}
    &H_{II} = H^0_{II} +\overbrace{H_{II}^{\delta_1} + H_{II}^{\epsilon_0} + H_{II}^{\epsilon_2}}^{=\,V} \\
    &H_{II}^0 = \frac{1}{2}[(\Omega_1-\widetilde{\Omega}_1)\sigma_x + \Omega_2\sigma_y] \\
    & H_{II}^{\delta_1} =  \frac{\delta}{2}[( \sin(\widetilde{\Omega}_1 t)\sigma_y + \cos(\widetilde{\Omega}_1 t)\sigma_z] \\
    & H_{II}^{\epsilon_0} = \frac{\epsilon}{2}[( \Omega_1\sigma_x + \frac{\Omega_2}{2}\sigma_y] \\
    & H_{II}^{\epsilon_2} = \frac{\epsilon}{2}\left[\frac{\Omega_2}{2}\cos(2\widetilde{\Omega}_1 t)\sigma_y - \frac{\Omega_2}{2}\sin(2\widetilde{\Omega}_1 t)\sigma_z\right].
\end{align}
\end{subequations}

As a first step, we determine the optimal modulation frequency $\widetilde{\Omega}^o_1$ by considering only the time-independent component of the Hamiltonian and disregarding the rotating noise contributions $H_{II}^{\delta_1}$ and $H_{II}^{\epsilon_2}$. Applying the optimization procedure from \cite{PhysRevLett.132.223601}, which is based on minimizing the variance of the energy gap, we find (to leading order in $\sigma_\epsilon$) that 
\begin{equation}\label{eq:optimal_pmdd_modulation}
    \widetilde{\Omega}^o_1 = \Omega_1 + \frac{1}{2}\frac{\Omega_2^2}{\Omega_1}.
\end{equation}
This is the expression employed in the quantum sensing and atomic clock examples in the Applications section. We shall adopt this choice henceforth, and we remark that, after this substitution, neglecting the rotating noise terms remains self-consistent, since they do not oscillate at the energy gap of $H_{II}^0$.

The next step is to derive an effective time-independent Hamiltonian and corresponding energy gap for $H_{II}$, and to compute its variance. This, in turn, enables the evaluation of a proxy for the coherence time via
$T_2 \propto \frac{\sqrt{2}}{\sqrt{\text{Var}(\Delta E)}}
\overset{\mathrm{def}}{=} \overline{T}_2$ \cite{PhysRevLett.132.223601}. At this point, all components of Eq. \eqref{eq:pmdd_subequations} are included, since under Eq. \eqref{eq:optimal_pmdd_modulation} all first-order contributions vanish (in the interaction picture with respect to $H_{II}^0$, no time-independent terms survive).

This is achieved using Floquet theory \cite{rudner2020floquetengineershandbook,leskes2010floquet,ivanov2021floquet}, made possible by the $\frac{2\pi}{\widetilde{\Omega}_1}$ time periodicity of $H_{II}$. Stroboscopically, Floquet states' dynamics is governed by such a Hamiltonian, for which we can find an energy gap through perturbation theory in $V$. We observe that, for our case, the Floquet-perturbation theory converges at the fourth order, which requires to truncate the infinite Floquet Hamiltonian at dimension 18. As a side note, we found that the necessary truncation dimension for order $K$ follows $dim^{Floquet}_{truncate}(K) = dim^H \cdot \bigl(1 + 2 \cdot h  \left\lfloor \frac{K}{2} \right\rfloor\bigr)$, where $h$ is the highest harmonic in the Floquet expansion of $H$, as may be expected. The full details of this derivation are provided in the SI.

The resulting analytical expression for $\overline{T}_2(\Omega_1,\Omega_2,\sigma_\epsilon)$ is numerically useful but algebraically quite involved and does not readily lend itself to a practical power series expansion in $\sigma_\epsilon$, which hinders further analytical progress. Nevertheless, we derived a practical approximate expression by observing that, at the optimum, $\overline{T}^o_2\propto \sigma_\epsilon^{-1}$, $\Omega^o_1 \propto \sigma_\epsilon^{-1/2}$, and $\Omega^o_2\propto \sigma_\epsilon^0$. This observation suggests a scaling ansatz, from which we obtain
\begin{equation}\label{eq:T2app}
\begin{split}
    &\overline{T}^{app}_2 
    \equiv 
    \frac{1}{\sigma_\epsilon} 
    \lim_{\sigma_\epsilon \to 0^+} 
    \left[\sigma_\epsilon \, \overline{T}_2\left(\frac{\Omega^s_1}{\sqrt{\sigma_\epsilon}},\Omega_2,\sigma_\epsilon\right)\right]
    \bigg|_{\Omega^s_1\,\mapsto\,\Omega_1\sqrt{\sigma_\epsilon}} \\
    &= 
    \frac{2 {T^\ast_2}^4 \Omega _1^2 \Omega _2}{\sqrt{4 {T^\ast_2}^4\Omega _2^4-24 {T^\ast_2}^2\Omega _2^2+{T^\ast_2}^8\Omega _1^8 \sigma _{\epsilon
   }^4+6 {T^\ast_2}^4 \Omega _1^4 \sigma _{\epsilon }^2+48}},
\end{split}
\end{equation}
where the $T^\ast_2$ scaling was reintroduced in the last line for completeness.

As we demonstrate below, this expression captures the qualitative behavior of $T_2$ in the sense that the two quantities are approximately monotonically related. Consequently, while we typically find that $1\lesssim \frac{T_2}{\overline{T}^{app}_2}\lesssim 2$, the $T_2$-maximizing control parameters ($\Omega_i$) can be obtained, to a good approximation, by maximizing $\overline{T}^{app}_2$. This is a remarkable outcome for a scaling ansatz approximation derived for an effective energy-gap variance that was obtained via Floquet perturbation theory, and its range of validity can be confirmed with case-specific numerical simulations.

\begin{figure}
    \centering
    \includegraphics[width=1\linewidth]{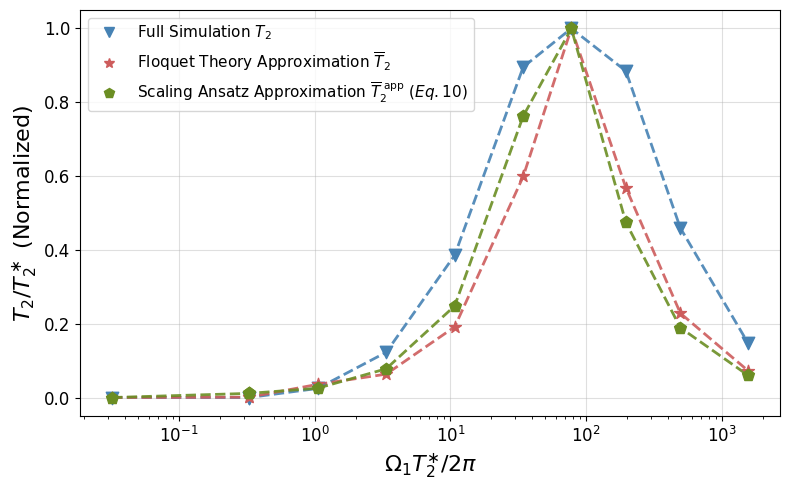}
    \caption{Coherence time as a function of $\Omega_1$ ($T^\ast_2$-scaled, normalized), demonstrating the validity of our approximation in Eq. \eqref{eq:T2app}. The full simulation results (blue traingles) compare well with the Floquet $4^{th}$-order perturbation theory derived $\overline{T}_2=\frac{\sqrt{2}}{\sqrt{Var(\Delta E)}}$ (red stars), and its scaling ansatz approximation $\overline{T}^{app}_2$ (green pentagons, Eq. \eqref{eq:T2app}). The optimization of our protocol's control parameters can, therefore, begin with our simple $\overline{T}^{app}_2$ formula, followed by more fine-grained numerical simulation and validation.}
    \label{fig:TheoryT2}
\end{figure}

In the atomic clock case, $T_2$ is optimized subject to the constraint given by the combined magic-angle and correlated-noise detuning conditions, resulting in $\Omega_2=\sqrt{2}\Omega_1$ (see Applications section). This can be directly substituted into Eq. \eqref{eq:T2app} to obtain the optimal Rabi $\Omega_1$. Figure \ref{fig:TheoryT2} shows that both $\overline{T}_2$ and $\overline{T}^{app}_2$ reproduce $T_2$ well, up to the aforementioned normalization factor, and give the correct $\Omega^{max}_1$.

Finally, in the absence of constraints on the control parameters, as in sensing or clock applications (e.g., quantum memory), we obtain the global optimum
\begin{equation}\label{eq:global_optimum}
    T^o_{2} \approx 1.5 \times \frac{T_2^\ast}{\sigma_{\epsilon}}, \
    \Omega^o_1 \approx 0.5 \times \frac{1}{T_2^\ast\sqrt{\sigma_{\epsilon}}}, \
    \Omega^o_2 \approx 0.5 \times \frac{1}{T_2^\ast};
\end{equation}
which demonstrates superior scaling compared to the single-drive case - 
$
    T^o_{2,sd} \approx 1.25 \times \frac{T_2^\ast}{\sqrt{\sigma_{\epsilon}}}
    , \ 
    \Omega^o_{1,sd} \approx 0.6 \times \frac{1}{T_2^\ast\sqrt{\sigma_{\epsilon}}}.
$
At this optimum, the amplitude ratio scales as $\sqrt{\sigma_\epsilon}$ and therefore, for typical values of amplitude noise, incorporating the Bloch–Siegert shift \cite{bloch1940magnetic, aharon2019quantum} often suffices to recover the optimal $T_2\propto T^\ast_2\sigma_\epsilon^{-1}$ scaling, as observed in other approaches \cite{PhysRevLett.132.223601,CaiNJP2012}. In contrast, we have shown that the same scaling can be maintained over a very wide bandwidth by eliminating the CRT of the control jointly with the destructive interference of correlated noise.

\section{Conclusion}\label{Section:Conclusion}
In this work, we have presented a control scheme that enables complete dynamic manipulation of the dressed qubit Hamiltonian while requiring only a minimal single-axis coupling in the lab frame. This approach permits the generation of circularly polarized drive fields in the dressed basis, removing the (sometimes energy-inefficient) dressed counter-rotating term, obviating the need for the dressed rotating-wave approximation, thus simplifying the dynamics in the strong driving regime ($\Omega_2\sim\Omega_1$).

As we demonstrate numerically, our scheme allows fast single qubit gates in the dressed basis, brings the two-qubit gate infidelity in a trapped-ion quantum processor an order-of-magnitude below the fault-tolerant threshold, enhances the sensitivity tenfold for high Rabi-range quantum sensing, thereby extending its range, and prolongs the coherence time for magic-angle constrained drive, as recently used for atomic clocks.

Our work can be extended in several key ways. 
First, control schemes involving dynamic modulation of the energy splitting $\Omega_1\to\Omega_1(t)$ (e.g. for electron-nuclear spin coupling \cite{casanova2019modulated,aharon2019quantum}, or global control in quantum dots \cite{SMARThuang2024high,SMARThansen2022implementation,SMARTPhysRevA.104.062415}) could potentially be combined with frequency modulated dressed drives $\widetilde{\Omega}_1\to\widetilde{\Omega}_1(t)$  to enhance performance. 
Second, our parametrization can be used as the starting point for more elaborate optimization methods, such as quantum optimal coherent control and pulse shaping. 
Third, our protocol can be examined for bare qubit control, potentially replacing the classic waveforms in certain scenarios. This includes, for example, creating a set of building-block pulses of high-fidelity for subsequent use in various pulse schemes. Forth, polarization transfer constitutes another interesting avenue, as similar schemes have been recently used in this context as well \cite{dagys2024robust,marshall2023radio,korzeczek2024towards,korzeczek2025phip}.

In summary, our work paves the way for a broader use of dressed qubits in various quantum technological application, especially where continuous operation is necessarily the case.

\begin{acknowledgments} 
\textit{Acknowledgments} --- 
A.S. thanks Daniel Cohen, Genko Genov, and Lennart Pelzer for fruitful discussions.
A.S. gratefully acknowledges the support of the Clore Israel Foundation Scholars Programme, the Israeli Council for Higher Education, and the Milner Foundation.
A. R. acknowledges the support of European Research
Council grant QRES, Project No. 770929, Quantera grant
MfQDS, Israel Science Foundation and the Schwartzmann
university chair.
\end{acknowledgments}

\bibliography{references}

\end{document}


\title{- Supplemental Information -\\ Optimally Driven Dressed Qubits}

\author{Alon Salhov \orcidlink{0000-0001-9111-2400}}
\email[Corresponding author: ]{alon.salhov@mail.huji.ac.il}
\affiliation{Racah Institute of Physics, The Hebrew University of Jerusalem, Jerusalem 9190401, Israel}
\author{Sagi Nechushtan}
\affiliation{Racah Institute of Physics, The Hebrew University of Jerusalem, Jerusalem 9190401, Israel}
\author{Alex Retzker}
\affiliation{Racah Institute of Physics, The Hebrew University of Jerusalem, Jerusalem 9190401, Israel}

\date{\today}

\maketitle



\tableofcontents

\setcounter{figure}{0}
\renewcommand{\thefigure}{S.\arabic{figure}}
\setcounter{equation}{0}
\renewcommand{\theequation}{S.\arabic{equation}}




\section{Dressed Qubit Control - Prior Art}
The first dressed-qubit control protocol that we survey, also called the second-order concatenated continuous drive \cite{CaiNJP2012,StarkNatComm2017}, sets (in the language of Eq.~(2) in the main text) $\vec{\Omega}(t) =(\Omega_1,2\Omega_2 \cos{(\Omega_1 t)},0)$, and is sometimes referred to as the double drive. Clearly, the $y$ component represents an orthogonal linearly polarized resonant drive of an $x$-basis dressed qubit. In analogy with bare qubits, we can transform the Hamiltonian to a second rotating frame with respect to $\frac{1}{2}\Omega_1\sigma_x$ to obtain $\frac{1}{2}\Omega_2\sigma_y$, which represents the control of the dressed qubit with second Rabi frequency $\Omega_2$.
However, this assumes that the dressed counter-rotating term (CRT) can be neglected, namely, it is valid under the dressed rotating wave approximation (RWA), $\Omega_2\ll\Omega_1$. The CRT limits, among other things, the speed of single-qubit gates (see further discussion in the Applications section and the SI section below). Moreover, half of the $y$-component amplitude $(\max|\Omega_y(t)| = 2*\Omega_2)$  is "wasted" on the CRT, which further limits performance when there are control amplitude or power constraints \cite{casanova2019modulated, aharon2019quantum}.

The second protocol \cite{CohenFP2017,farfurnik2017experimental}, sets $\vec{\Omega}(t) =(\Omega_1,0,-2\Omega_2 \cos{(\Omega_1 t))}$, by using a time-dependent detuning (frequency/phase modulation). This scheme is preferable when the control field's phase stability exceeds its amplitude stability, which is often the case. Leaving this difference aside, the changed drive component ($y\rightarrow z$) and the added minus sign are immaterial when it comes to the CRT, and all the comments from the first protocol hold here as well.

The third protocol we review is Mixed Dynamical Decoupling \cite{Genov2019MDD}, where a continuous first drive is mixed with a pulsed second drive. By setting $\vec{\Omega}(t) =(\Omega_1,-2\Omega_2*(\sum_{i=1}^n \operatorname{rect}(\frac{t - t_i}{\tau_i}) \cos{(\Omega_1 t +\phi_i),0)}$, the dressed qubit is controlled by a sequence of rectangular pulses of duration $\tau_i$, flip-angles $\Omega_2\tau_i$, and phases $\phi_i$, at times $t_i$. Once more, the dressed RWA ($\Omega_2\ll\Omega_1)$ must be satisfied, severely limiting the manipulation speed.

More recently, some of the authors demonstrated that destructive interference of cross-correlated control amplitude fluctuations ($\delta\Omega_i$) can be achieved by setting $\vec{\Omega}(t) =(\Omega_1,2\Omega_2 \cos(\Omega_1+(c+\frac{1}{4})\frac{\Omega_2^2}{\Omega_1}),0)$, where c is the cross-correlation \cite{PhysRevLett.132.223601}. Although the scheme significantly improves key metrics such as coherence time and sensitivity, and despite the fact that the Bloch-Siegert shift is integrated into the scheme (the $\frac{1}{4}$ factor),  the dressed RWA ($\Omega_2\ll\Omega_1$) still limits its performance, e.g., its sensing bandwidth.

These examples illustrate how the dressed RWA is typically assumed in all dressed qubit control protocols. Our control scheme eliminates this problem by creating a circular drive in the dressed basis, eliminating the CRT.

\newpage
\section{Single-Qubit Gate CRT-induced Infidelity}
\begin{figure}[h]
    \centering
    \includegraphics[width=0.5\linewidth]{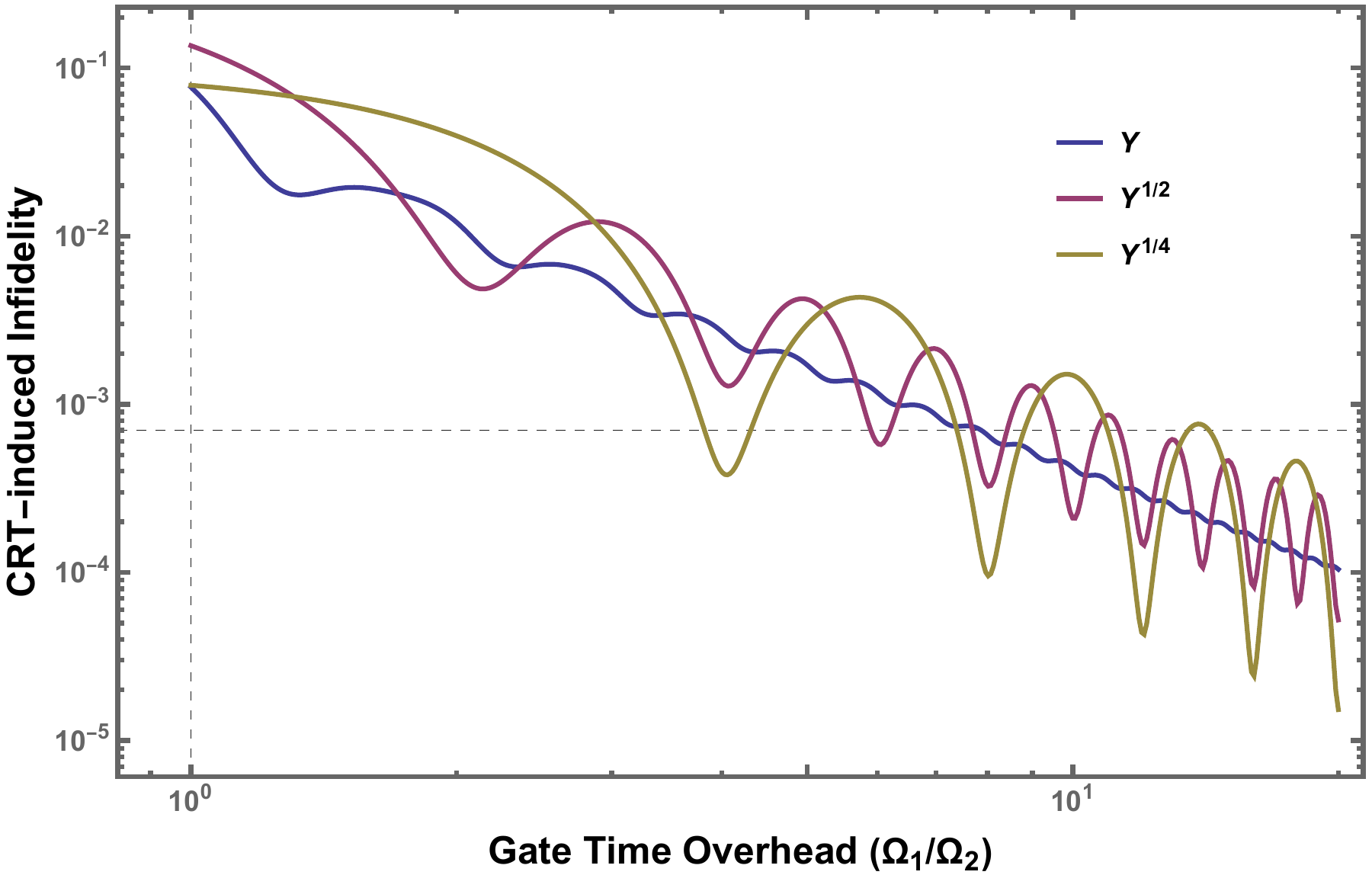}
    \caption{CRT-induced infidelity vs. the bare-dressed qubit gate overhead time $\Omega_1/\Omega_2$ for the double-drive scheme. To obtain sub-threshold infidelity ($< 7*10^{-3}$), the needed overhead time is $\gtrsim 10$. With our new scheme, as a result of the elimination of the CRT, no overhead time is required.}
    \label{fig:Fidelity1QG}
\end{figure}

In this section, we quantify the effect of the CRT on single qubit gate infidelities in the dressed basis, and the resulting required overhead time (compared to bare qubits). As representative examples, we consider the gate infidelities \cite{nielsen2002simple} of the $Y^{1/n}$ gates, for $n=1,2,4$, using the second-order concatenated continuous drive \cite{CaiNJP2012}. As mentioned in the previous section, this scheme is obtained by setting (in the language of Eq.~(2) in the main text) $\vec{\Omega}(t) =(\Omega_1,2\Omega_2 \cos{(\Omega_1 t)},0)$. Figure \ref{fig:Fidelity1QG} shows the CRT-induced infidelity (which is the only source of infidelity in this case) vs. the bare-dressed qubit gate overhead time $\Omega_1/\Omega_2$ (see the main text for further context). To obtain sub-threshold infidelity, the overhead time should be at least an order-of-magnitude. As mentioned in the main text, this issue is completely avoided using our new control protocol. The CRT frequency ($2\Omega_1$) manifests itself in the infidelity oscillations as a function of $\Omega_1/\Omega_2$. The effect depends on the phase acquired during the gate time ($t_g$), namely $2\Omega_1 *t_g=2\Omega_1 * \frac{\pi/n}{\Omega_2}=2\pi\frac{1}{n}*\frac{\Omega_1}{\Omega_2}$, exhibiting $1/n$ oscillations.

\newpage
\section{Two-Qubit Gate Fidelity}
The Hamiltonian describing the system of two ionic qubits and a motional mode, modeled as an harmonic oscillator (HO), appears in Eq.~(4) in the main text. As in \cite{PhysRevX2025Daniel}, by transforming to the interaction picture with respect to
$\sum_{j=1,2} \frac{1}{2}(\omega_0^{(j)} +\Omega_2\cos(\Omega_1 t)) \sigma_z^{(j)}$ we obtain:

\begin{equation}\label{eq:ion_full}
    H_I(t) = \nu b^\dagger b + \sum_{j=1,2}( \frac{\eta\nu}{2}\sigma_z^{(j)}(b+b^\dagger) +H_{D,I}^{(j)}(t)),
\end{equation}
where the parameters $\nu=2\pi\times98.8$ kHz, $\eta=0.033$ are set according to their value in \cite{PhysRevX2025Daniel}. The drive Hamiltonians in this picture are given by:
\begin{equation}\label{eq:ion_drive}
    H_{D,I}^{(j)}(t) = 
    \frac{\Omega_1}{2}\sigma_x^{(j)}
    -\frac{\Omega_2}{2}\cos(\Omega_1t)\sigma_z^{(j)}
    +\ s \ * \ \frac{\Omega_2}{2}\sin(\Omega_1t)\sigma_y^{(j)},
\end{equation}
where we set $\Omega_1 = \nu-\eta\nu$ (as in \cite{PhysRevX2025Daniel}), and $\Omega_2$ is the parameter that we optimize as discussed in the main text. The parameter $s$ distinguishes between the scheme in \cite{PhysRevX2025Daniel}, which has $s=0$ and therefore suffers from the CRT (see main text), and our new scheme which does not, by having $s=1$.

The final two-qubit density matrix, after the entangling operation $U = \mathcal{T} \exp\left(-i\int_{0}^{t_g} H_I(t') \, dt'\right)$ for a gate time $t_g\approx \frac{2\pi}{\eta\nu}$ (the precise gate time is optimized so that the two-qubit purity is maximal), is given by $\rho = \text{Tr}_{\text{HO}} \left[ U \left( \rho_i \otimes \rho^{\text{th}}_{HO} \right) U^\dagger \right]$, where $\rho_i=|+,+\rangle\langle+,+|$ is the initial two-qubit density matrix (representing a product state) and $\rho^{\text{th}}_{HO}(\bar n)$ is the initial density matrix of the motional mode, which starts in a thermal state with average phonon number $\bar n$. Following the experiment in \cite{PhysRevX2025Daniel}, we set $\bar n = 0.6$.

Following \cite{PhysRevX2025Daniel}, we define the fidelity of the entanglement operation as
\begin{equation}
    \EuScript{F} = \underset{R_j}{\text{max}}\langle \Phi^+ | R_1^\dagger  R_2^\dagger \rho  R_1  R_2 | \Phi^+ \rangle,
\end{equation}
where $|\Phi^+ \rangle= (|00 \rangle + |11 \rangle)/\sqrt{2}$
is the target Bell-state of the entangling operation and $R_{j}$ is a single-qubit rotation of qubit $j$. The fidelity is defined by optimizing over single-qubit gates.

\newpage
\section{Coherence Time Simulation}
As discussed in the main text, to evaluate the coherence time we employ a noise model with two contributions, which are often the limiting factors for dressed qubit performance. The first arises from environmental fluctuations of the bare qubit energy gap, modeled as $\omega_0 \to \omega_0 + \delta$ (in Eq.~(1)), with $\delta \sim \mathcal{N}\!\left(0,\; \frac{2}{T_2^{\ast2}}\right)$ and $T_2^\ast$ denoting the bare qubit’s Ramsey dephasing time. The second contribution stems from control noise, described as fractional fluctuations in the drive amplitude, modeled by $f(t) \to f(t)(1+\epsilon)$, where $\epsilon \sim \mathcal{N}\!\left(0,\;\sigma_\epsilon^2\right)$.

Note that this assumes a perfect correlation between the amplitude fluctuations of the two amplitudes $\delta\Omega_{x/y}$, as recently demonstrated in \cite{PhysRevLett.132.223601}, and the high stability of the drive's frequency and phase control. The noise terms are modeled as random variables, not processes, effectively assuming that their correlation time is longer than the time of a single experiment. Note, however, that decoupling is often effective for shorter correlation times, as long as they are longer than the typical control timescale, or Rabi period, as demonstrated in \cite{PhysRevLett.132.223601}.

For each control scheme specified by $\vec\Omega(t)$, we numerically simulate the dynamics governed by Eq.~(2) in the main text, incorporating the replacements prescribed by the noise model. In the first interaction picture with respect to $\frac{1}{2}(\omega_0-\Omega_z(t))\sigma_z$, the Hamiltonian takes the form
\begin{equation}\label{SIEq:noisy_rotating_frame_Hamiltonian}
       H_I=
       \frac{1}{2}[\Omega_x(t)(1+\epsilon)\sigma_x + \Omega_y(t)(1+\epsilon)\sigma_y + (\Omega_z(t)+\delta)\sigma_z] = 
       \frac{1}{2}\vec{\Omega}(t)\cdot\vec{\sigma}
       +
       \frac{1}{2}
            \begin{pmatrix}
            \Omega_x(t)\epsilon \\
            \Omega_y(t)\epsilon \\
            \delta
            \end{pmatrix}
       \cdot\vec{\sigma}
       ,
\end{equation}
from which we obtain the propagator for a single $(\delta,\epsilon)_n$ noise realization as $U_n(t)= \mathcal{T} \exp\left(-i\int_{0}^{t} H_{I,n}(t') \, dt'\right)$. The density matrix then evolves according to $\rho_{k,n}(t)=U_n(t)\,\rho_k(0)\,U_n^\dagger(t)$, where $\rho_k(0) = \frac{1}{2}(\sigma_0+\sigma_k),\ k=x,y,z$ denotes the initial ($t=0$) density matrices corresponding to the positive-eigenvalue eigenstate of the Pauli operator $\sigma_k$. We exploit the time-periodicity of the Hamiltonian and compute the propagator using QuTiP’s Floquet solver implementation~\cite{qutip5}.

We generate a set of $N=2048$ noise realizations $(\delta,\epsilon)^{N}_{n=1}$ using a scrambled 2D Sobol sequence, selected for its favorable numerical characteristics. The corresponding expected density matrix is obtained by averaging over all realizations $\bar\rho_k(t)=\frac{1}{N}\sum_{n=1}^N\rho_{k,n}(t)$. The quantum memory fidelity is defined as $F(t)=\frac{1}{3}\sum_{k=x,y,z}\operatorname{Tr}(\bar\rho_k(t)\rho_k(0))$ \cite{BowdreyPhLett2002}, and the coherence time $T_2$ is extracted as the time at which the fidelity envelope decreases from $1$ to $\frac{2+1/e}{3}\approx0.79$. This threshold corresponds to a $1/e$ decay of the fastest-dephasing (worst-case, least protected) states \cite{PhysRevLett.132.223601}.

\newpage
\section{Floquet Perturbation Theory -  Effective Energy Gap Variance}

In the main text, we discuss a derivation of an effective time-independent Hamiltonian for $H_{II}$ (Eq.~8), along with its corresponding energy gap ($\Delta E$). This gap is then employed to define a proxy for the coherence time,
$
\overline{T}_2
\overset{\mathrm{def}}{=} 
\frac{\sqrt{2}}{\sqrt{\text{Var}(\Delta E)}}
$ \cite{PhysRevLett.132.223601},
which is subsequently used to extract a simple formula based on a scaling ansatz. In this section, we present the detailed derivation of $\overline{T}_2$. In the following analysis, we do not employ any further interaction-picture transformations, so we will drop the ${II}$ subscript from this point onward. Furthermore, as stated in the main text, we keep $\widetilde{\Omega}_1 = \Omega_1 + \frac{1}{2}\frac{\Omega_2^2}{\Omega_1}$ (Eq.~9) fixed throughout this section.

We decompose $H$ into its unperturbed and time-independent part $H^{(0)}$, and the time-dependent and $\frac{2\pi}{\widetilde{\Omega}_1}$-periodic perturbation $V$. The latter is further decomposed into its Fourier components $V^{(m)}$ as follows:

\begin{subequations}\label{eq:pmdd_subequations}
\begin{align}
    &H = H^0 +\overbrace{H^{\delta_1} + H^{\epsilon_0} + H^{\epsilon_2}}^{=\,V} = H^{(0)}+\sum_{m=-2}^2 e^{-im\widetilde{\Omega}_1 t} V^{(m)} \\
    &H^0 = \frac{1}{2}[(\Omega_1-\widetilde{\Omega}_1)\sigma_x + \Omega_2\sigma_y] = H^{(0)}\\
    & H^{\delta_1} =  \frac{\delta}{2}[( \sin(\widetilde{\Omega}_1 t)\sigma_y + \cos(\widetilde{\Omega}_1 t)\sigma_z] = e^{-i\widetilde{\Omega}_1 t} V^{(1)} + e^{+i\widetilde{\Omega}_1 t} V^{(-1)} \\
    & H^{\epsilon_0} = \frac{\epsilon}{2}[( \Omega_1\sigma_x + \frac{\Omega_2}{2}\sigma_y] = V^{(0)} \\
    & H^{\epsilon_2} = \frac{\epsilon}{2}\left[\frac{\Omega_2}{2}\cos(2\widetilde{\Omega}_1 t)\sigma_y - \frac{\Omega_2}{2}\sin(2\widetilde{\Omega}_1 t)\sigma_z\right] = e^{-i2\widetilde{\Omega}_1 t} V^{(2)} + e^{+i2\widetilde{\Omega}_1 t} V^{(-2)}.
\end{align}
\end{subequations}
We then diagonalize $H^{(0)}$ using 
$$
U_0
= \frac{1}{\sqrt{2}}
\left(
\begin{array}{cc}
 \frac{z}{| z| } & -\frac{z}{| z| } \\
 1 & 1 \\
\end{array}
\right), z=(\widetilde{\Omega}_1-\Omega _1)+i \Omega _2;
$$ 
such that $H^{\prime(0)} = U_0^\dagger H^{(0)} U_0 = -\frac{1}{2} |z| \sigma_z$, and apply the same transformation to the perturbation $V^{\prime (m)}= U_0^\dagger V^{(m)} U_0$.

We can now formulate the effective time-independent Hamiltonian in the extended Floquet space \cite{rudner2020floquetengineershandbook,leskes2010floquet,ivanov2021floquet}. Although this space is, in principle, infinite-dimensional, it can be truncated to a finite size that depends on the required perturbation order $K$. We find that the truncation dimension is given by $dim^{Floquet}_{truncate}(K) = dim^H \cdot \bigl(1 + 2 \cdot h  \left\lfloor \frac{K}{2} \right\rfloor\bigr)$, where $h$ denotes the highest harmonic included in the Floquet expansion of $H$, as may be expected. In addition, we observe that $\overline{T}_2$ converges for $K=4$ (in particular, we focus on the region near its maximum). Substituting the relevant values for our system ($dim^H=2, h=2, K=4$), we obtain $dim^{Floquet}_{truncate}=18$.

The effective Floquet Hamiltonian is given, therefore, by $H_{F} = $
\begin{equation}
\resizebox{1.0\textwidth}{!}{$
\setcounter{MaxMatrixCols}{20}
\begin{pmatrix}
H^{\prime(0)} + V^{\prime(0)} + 4\tilde{\Omega}_1 & V^{\prime(-1)} & V^{\prime(-2)} & 0 & 0 & 0 & 0 & 0 & 0 \\
V^{\prime(1)} & H^{\prime(0)} + V^{\prime(0)} + 3\tilde{\Omega}_1 & V^{\prime(-1)} & V^{\prime(-2)} & 0 & 0 & 0 & 0 & 0 \\
V^{\prime(2)} & V^{\prime(1)} & H^{\prime(0)} + V^{\prime(0)} + 2\tilde{\Omega}_1 & V^{\prime(-1)} & V^{\prime(-2)} & 0 & 0 & 0 & 0 \\
0 & V^{\prime(2)} & V^{\prime(1)} & H^{\prime(0)} + V^{\prime(0)} + \tilde{\Omega}_1 & V^{\prime(-1)} & V^{\prime(-2)} & 0 & 0 & 0 \\
0 & 0 & V^{\prime(2)} & V^{\prime(1)} & H^{\prime(0)} + V^{\prime(0)} & V^{\prime(-1)} & V^{\prime(-2)} & 0 & 0 \\
0 & 0 & 0 & V^{\prime(2)} & V^{\prime(1)} & H^{\prime(0)} + V^{\prime(0)} - \tilde{\Omega}_1 & V^{\prime(-1)} & V^{\prime(-2)} & 0 \\
0 & 0 & 0 & 0 & V^{\prime(2)} & V^{\prime(1)} & H^{\prime(0)} + V^{\prime(0)} - 2\tilde{\Omega}_1 & V^{\prime(-1)} & V^{\prime(-2)} \\
0 & 0 & 0 & 0 & 0 & V^{\prime(2)} & V^{\prime(1)} & H^{\prime(0)} + V^{\prime(0)} - 3\tilde{\Omega}_1 & V^{\prime(-1)} \\
0 & 0 & 0 & 0 & 0 & 0 & V^{\prime(2)} & V^{\prime(1)} & H^{\prime(0)} + V^{\prime(0)} - 4\tilde{\Omega}_1
\end{pmatrix}
$},
\end{equation}
with the diagonal part $H_F^{(0)} = H_F\big|_{V^{\prime (m)} \to 0}\overset{
}{=}E_n^{(0)}\ket{n^{(0)}}\bra{n^{(0)}}$
, the effective perturbation $V_F = H_F-H_F^{(0)}$, and the effective energy gap which is by the difference between the middle energies $\Delta E = E_{9}-E_{10}$.

The energies $E_n$ are calculated using perturbation theory, to forth order. 
Denoting $V_{nm} = \bra{n^{(0)}}V_F\ket{m^{(0)}}$ and $E_{nm} = E_n^{(0)}-E_m^{(0)}$, these are given by:

\begin{subequations}
\begin{align}
    & E_n^{K} = \sum_{k=0}^{K=4} E_n^{(k)} \\
    & E_n^{(0)} =  \bra{n^{(0)}}H_F^{(0)}\ket{n^{(0)}} \\
    & E_n^{(1)} =  V_{nn}
    \\
    & E_n^{(2)} = \sum_{m_1 \neq n} \frac{| V_{nm_1} |^2}{E_{nm_1}}
    \\
    & E_n^{(3)} = \sum_{m_1,m_2 \neq n}
    \frac{V_{nm_1} V_{m_1 m_2} V_{m_2n}}
    {E_{nm_1}E_{nm_2}} 
    - V_{nn} \sum_{m_1 \neq n} 
    \frac{|V_{nm_1}|^2}{E_{nm_1}^2} \\
    & E_n^{(4)} = 
    \sum_{m_1,m_2,m_3 \neq n}
    \frac{V_{nm_1}V_{m_1m_2}V_{m_2m_3}V_{m_3n}}{E_{nm_1}E_{nm_1}E_{nm_1}} 
    - E^{(2)}_n \sum_{m_1 \neq n}
     \frac{|V_{nm_1}|^2}{E_{nm_1}^2} 
    - 2V_{nn}\sum_{m_1,m_2 \neq n}
    \frac{V_{nm_1}V_{m_1m_2}V_{m_2n}}{E_{nm_1}E_{nm_2}^2} 
    + V_{nn}^2 \sum_{m_1 \neq n} \frac{|V_{nm_1}|^2}{E_{nm_1}^3}.
\end{align}
\end{subequations}

Our energy gap then reads $\Delta E = E_{9}^K-E_{10}^K$ and, as previously defined, $\overline{T}_2 = \frac{\sqrt{2}}{\sqrt{Var(\Delta E)}}$.
The resulting analytical expression is numerically useful but algebraically quite involved and does not readily lend itself to a practical power series expansion in $\sigma_\epsilon$, which hinders further analytical progress. As mentioned in the main text, we obtain a practical analytical expression using the scaling ansatz.

\bibliography{references}